# Experimental Analysis of Subscribers' Privacy Exposure by LTE Paging

Christian Sørseth · Xianyu Shelley Zhou · Stig F. Mjølsnes · Ruxandra F. Olimid

**Abstract** Over the last years, considerable attention has been given to the privacy of individuals in wireless environments. Although significantly improved over the previous generations of mobile networks, LTE still exposes vulnerabilities that attackers can exploit. This might be the case of paging messages, wake-up notifications that target specific subscribers, and that are broadcasted in clear over the radio interface. If they are not properly implemented, paging messages can expose the identity of subscribers and furthermore provide information about their location. It is therefore important that mobile network operators comply with the recommendations and implement the appropriate mechanisms to mitigate attacks. In this paper, we verify by experiment that paging messages can be captured and decoded by using minimal technical skills and publicly available tools. Moreover, we present a general experimental method to test privacy exposure by LTE paging messages, and we conduct a case study on three different LTE mobile operators.

Christian Sørseth, Xianyu Shelley Zhou, Stig F. Mjølsnes and Ruxandra F. Olimid
Department of Information Security and Communication Technology, NTNU, Norwegian University of Science and Technology, Trondheim, Norway
E-mail: sfm@ntnu.no,ruxandra.olimid@ntnu.no
Note: Christian Sørseth and Xianyu Shelley Zhou performed most of the work during their MSc studies at NTNU.

Ruxandra F. Olimid
Department of Computer Science, University of Bucharest, Romania
E-mail: ruxandra.olimid@fmi.unibuc.ro

## 1 Introduction

Long-Term Evolution (LTE) is the fastest developing mobile network technology of all times, commercially launched by more than 670 operators worldwide, and with an estimation of up to 700 until the end of 2018 [14]. In these circumstances, disclosure of private information in LTE brings important consequences, both for 4G as a standalone technology, but also because 4G is a main pillar for the coming 5G [16,32,31]. Although it is significantly improved over the previous generation networks, LTE still exposes vulnerabilities that can easily be exploited by an adversary to break the privacy of subscribers. Examples include IMSI catchers, active devices that masquerade genuine base stations with the purpose of stealing the permanent identity of subscribers. This makes LTE vulnerable to location attacks such as testing the presence or absence of a subscriber in an area [33,17,24,25]. Another example is the one we address in this paper: privacy exposure of subscribers by paging messages. Paging messages are wake-up messages sent by the network to a device to notify it of an event, such as a call or data reception. They are broadcast messages targeting a specific subscriber, so they need to contain a sort of identification. Because paging messages are broadcasted prior to the establishment of a secure communication channel, the identification string they contain is sent in clear. Assuming implicit trust in pre-secured communication is a problem in nowadays mobile networks [16,32]. Hence, to some extent, paging messages can expose private information. Unlike IMSI catching, paging related attacks are mostly passive, so they are more difficult to detect and mitigate. The current 3rd Generation Partnership Project (3GPP) standardisation fails to implement countermeasures for paging attacks in LTE, and so there is a strong motivation to study of paging mes-sages and related attacks. Moreover, by exploiting identy leakage, paging attacks relate to a well known issue in mobile communications, *the private identification problem*, or how to securely send the identification string over the network so that it remains secure against unauthorised parties [23]. Motivated by all the above-mentioned aspects, the paper presents an experimental method to test at what extent the paging messages expose the privacy of subscribers, and propose a case study on three mobile operators.

### 1.1 Related Work

Shaik et al. proposed a solution to intercept LTE broadcast channels, collect and decode paging messages [33]. The implementation heavily relies on a Universal Software Radio Peripheral (USRP) and the open source software srsLTE, and it exploits specification and implementation vulnerabilities in LTE. Additionally, they explained semi-passive attacks for triggering paging messages to a specific subscriber by using social applications such as Facebook and WhatsApp. Similar work on analysing paging messages in commercial networks was done in the past for GSM networks [18].

Security vulnerabilities introduced by paging messages in LTE have been signalled many times in the literature [38,37,36,17,19,15,9]. Paging messages can be used to trigger Denial-of-Service (DoS) attacks by simply planting a phone in a targeted area to respond all paging messages and ignore further communication [38]. Another way to induce DoS attacks is by jamming, and work has been done to analyse and diminish the problem [37,19]. LTE has also been shown vulnerable to IMSI disclosure and location exposure by paging [36,9]. Solutions that claim to enhance the privacy of identifiers in paging messages and mitigate location disclosure have been proposed [15,36]. Applications that detect threats in mobile networks are currently available online, but they mostly do not support paging detection. Park et al. conducted an analysis for GSM and UMTS and found one Android application that monitors paging messages and triggers an alert when the paging message contains the permanent identifier of the subscriber (IMSI) [30]. Bojic et al. give a brief overview on the evolution of IMSI paging attacks from 2G/GSM to 4G/LTE [9].

### 1.2 Contribution

The vulnerabilities exploited in the last years show that there is still much space left for improvements in LTE with respect to privacy. This paper focuses on privacy exposure caused by LTE paging, and extends the existing work in the sense that it describes in detail an experimental method to test the privacy of subscribers in LTE networks. The experiments are all feasible with open source software, publicly available low-cost hardware and basic technical skills. They consist in capturing and analysing paging messages with respect to the identity of subscribers. Motivated by Shaik et al.'s remark that some network operators tend to not change the temporary identifiers [33], we also investigate how the temporary identifiers are used and refreshed. Finally, we introduce a method to verify if smart paging is implemented in an LTE network. All experiments are explained in sufficient detail, so that the procedures can be easily replicated.

Following the experimental method described, we conduct an experimental study on three different mobile operators in Norway: Telenor, Telia and ice.net. We present the results, analyse the data and discuss the findings. Overall, the paper pertains to LTE in general. The choice of election for the operators is only given by the geographical location of the authors.

### 1.3 Outline

The paper is organised as follows. Next section gives the necessary background on LTE. Section 3 introduces an experimental method for testing the privacy of subscribers' identity in LTE mobile networks. Section 4 presents and discusses the results of an experimental study conducted for the three Norwegian mobile operators. Last section concludes.

## 2 Background

3GPP developed LTE with the aim to increase downlink and uplink peak data rates, create scalable carrier bandwidths, and make a purely Internet Protocol (IP) based network architecture [1]. In addition to the significant functionality improvements, the security architecture has also been enhanced in comparison to its predecessors. This section reviews the overall LTE architecture, with focus on the subscribers' identifiers and paging messages that are directly used in our work.

### 2.1 LTE Architecture

The LTE network architecture is roughly divided into the access part called the Evolved Universal Terrestrial Radio Access Network (E-UTRAN) and the core part called the Evolved Packet Core (EPC). The E-UTRAN and EPC are themselves divided into several network components, each playing an important role in the complete LTE network architecture. Figure 1 illustrates the LTE architecture. We briefly summarise the network components next.

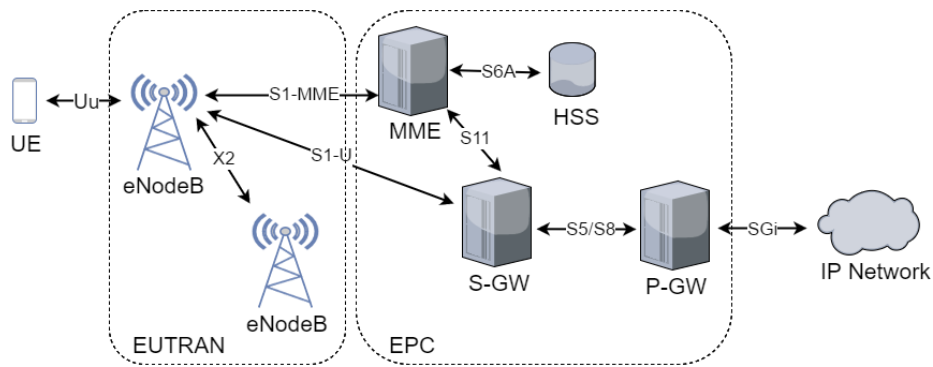

**Fig. 1** LTE network architecture [24]

*User Equipment (UE)* is the mobile terminal of the user. It consists of the Mobile Equipment (ME) such as a phone or a tablet, equipped with a card provided by the network operator, called Universal Subscriber Identity Module (USIM)[1].

*Evolved Node B (eNodeB)* is the base station responsible for the LTE radio functionality, being the access point of the UE to the network. Each eNodeB serves a coverage area, which is divided into several sectors known as *cells* [33]. Multiple cells group in Tracking Areas (TAs), with a TA being managed by a single MME.

*Mobility Management Entity (MME)* is a key control plane entity within the EPC, providing an interface towards the E-UTRAN. The primary responsibility of the MME is to manage the accessibility of network connections, allocate network resources, and authenticate UEs [34].

*Home Subscriber Server (HSS)* is essentially a database containing subscriber-related information such as the International Mobile Subscriber Identity (IMSI), authentication key, quality of service profile, and roaming restrictions [13,20]. The identities will be explained in more detail in the next section.

*PDN Gateway (P-GW)* is the interconnection node with external packet data networks such as the Internet.

*Serving Gateway (S-GW)* is the interconnection node between the EPC and the E-UTRAN anchor point for intra-LTE mobility.

## 2.2 LTE Identifiers

To access mobile services, a UE must successfully authenticate itself to the network by following an Authentication and Key Agreement (AKA) procedure. AKA is mainly based on a static shared identification string and a cryptographic key that are unique for each subscriber and are stored in the USIM at the user side and in the HSS at the network side. The permanent identification is called International Mobile Subscriber Identity (IMSI).

To maintain the privacy of the subscribers, the IMSI should be exposed as rarely as possible. To fulfill this, mechanisms to protect the permanent identity by allocating temporary identities are used. Global Unique Temporary Identity (GUTI) is a unique temporary identity allocated to the UE by the MME. GUTI identification is unambiguous and prevents the disclosure of the permanent identity IMSI at eavesdropping. The GUTI update interval is operator-specific and may vary among network operators. Figure 2 shows the complete structure of the GUTI [2]. The last part of the GUTI forms the SAE-TMSI (S-TMSI), which identifies the MME in the network by the MME Code (MMEC) and the UE by the MME Temporary Subscriber Identifier (M-TMSI). For UEs that camp in cells under the same MME, MMEC is the same, and only the M-TMSI differs. The M-TMSI, or equivalently the more complete identification S-TMSI, are important for our work, because they temporarily identify a subscriber. The M-TMSI is also stored (and refreshed) at both ends (user

---

| GUTI (Global Unique Temporary UE Identity) | | | | |
|---|---|---|---|---|
| GUMMEI (Globally Unique MME Identifier) | | | | M-TMSI (MME Temporary Subscriber Identifier) |
| MCC (Mobile Country Code) | MNC (Mobile Network Code) | MMEI (MME Identifier) | | |
| | | MMEGI (MME Group ID) | MMEC (MME Code) | |
| | | S-TMSI (SAE-TMSI) | | |

**Fig. 2** The structure of GUTI

side and visiting network side, the network to which the UE camps on). GUTI also contains the Mobile Country Code (MCC) and Mobile Network Code (MNC) that together uniquely identify the mobile operator worldwide, and the MME Group ID (MMEGI), which is an identifier for a group of MMEs in the network. They all compose the *Globally Unique MME Identifier (GUMMEI)*, which fully identifies the MME that allocated the GUTI.

### 2.3 LTE Paging

Paging is a Radio Resource Control (RRC) procedure used by LTE networks to notify one or more UEs of an event [4]. UEs are monitoring the paging channel to detect incoming calls, system information change, or other notifications such as for example Earthquake and Tsunami Warning System (ETWS) [5]. The paging messages identify the UEs by their temporary identity S-TMSI, which populates the `ue-Identity` field. Only the addressed UEs respond to the paging message. To avoid network congestion and increase efficiency, the operators page multiple UEs in a single message whenever this is possible. In this case, a single paging message contains more *paging records*, one for each paged UE. Figure 6(b) shows an example of a paging message decoded in xml format with a single paging record. Similarly, `systemInfoModification` is a field indicating that system information changes at the next modification period boundary, while `etws-Indication` indicates a ETWS notification [5].

Traditionally, a paging message is broadcasted in all cells of a TA. With *smart paging* enabled, the network pages a UE only in the last cell on which the UE camped and maybe the neighbour cells [29]. Only if the UE does not reply, the paging message is afterwards broadcasted in the whole TA. Even if it is not completely certain that a paged UE is indeed located in the paged cell, the probability is high, so smart paging decreases the amount of paging messages sent in the network and hence increases performance. On the other hand, it makes location tracking by eavesdropping more accurate. With smart paging enabled, a subscriber can be located within the coverage area of one cell, typically of two km², instead of the whole TA which is typically 100 km² [33].

### 3 Experimental Method

This section presents an experimental method for testing the privacy of subscribers' in commercial networks. All experiments are feasible to implement with open source software and publicly available hardware, and they can be fully reproduced following the described methodology.

### 3.1 Tools

We start by describing the software and hardware tools necessary to conduct the experiments.

#### 3.1.1 srsLTE

srsLTE is an open source platform developed by Software Radio Systems for configuring Software Defined Radio (SDR) network components such as UEs and eNodeBs [35]. The software is implemented in C, and it is compatible with the LTE Release 8 specification, defined by 3GPP [35,3]. srsLTE contains multiple prebuilt example applications based on UE and eNodeB specifications. Implemented applications of interest for our work are *cell_search*, which provides a list of neighbouring LTE cells together with their frequency and *pdsch_ue*, which behaves as a UE and provides measurement data of commercial LTE cells, catches and decodes paging messages.

#### 3.1.2 Ettus B200mini

Ettus B200mini is a Universal Software Radio Peripheral (USRP) manufactured by Ettus Research, intended to be a low-priced device for universities, research labs, and hobbyists [27]. The B200mini utilizes a wide frequency range (70MHz - 6GHz), which allows operation on all frequency bands in GSM, UMTS, and LTE [11]. B200mini's small size makes it portable and physically hard to detect. It is conveniently bus-powered by USB 3.0 connectivity, which allows it to be operated from anywhere with the help of a portable battery pack.

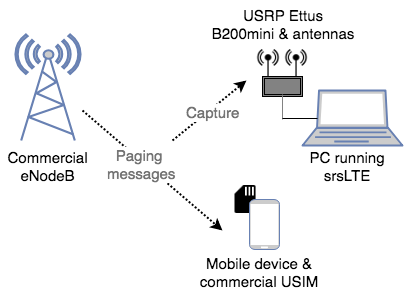

**Fig. 3** Experimental architecture

## 3.2 Experimental Setup

The experimental setup consists of a personal computer running srsLTE, attached with SDR and antennas. srsLTE software runs on the computer to collect the paging messages broadcasted by a commercial eNodeB on a given frequency. For our experiments, we used a desktop computer running Ubuntu 14.04 LTS, version 3.19.0-031900-low latency kernel, Inter Core i7 and Ettus B200mini USRP equipped with Pulse Electronics W1900 antennas [8]. The Ettus B200 mini USRP board only, respectively the board equipped with enclosure (manufactured locally at NTNU) and antennas is given in Figure 4. For testing the persistence of temporary identities, we used an iPhone, running iOS version 10 equipped with commercial USIMs belonging to the targeted commercial networks. Any equipment that accepts the commercial USIM and has the facility of displaying the M-TMSI can be used with success. Figure 3 illustrates the schematic experimental architecture.

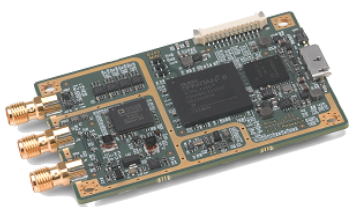

(a) USRP B200mini [11]

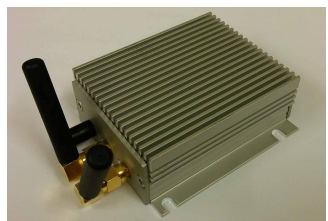

(b) USRP B200mini equipped with enclosure and antennas

**Fig. 4** USRP equipment

## 3.3 Methodology

The current subsection describes the methodology used to conduct the experiments. Figure 5 summarises the major steps, which are described in more details next. The lower branch (A-D) is only used for testing the persistence of temporary identities, while all the other experiments are following the upper branch.

### A. Experimental setup

Experimental setup consists in the acquisition of necessary hardware, and the installation and configuration of the software. The experimental architecture is illustrated in Figure 3 and explained in the previous subsection. More information about the tools and equipment was given in Subsection 3.1.

### B. Data capture

To proceed with capturing the data after the experimental setup is complete, the frequency of interest must be determined. This is a downlink frequency used by the target network to send paging messages to UEs in the area of the experiment.

### B1. Determine cell frequency

There exist multiple ways to determine the downlink frequency of an operator in the measurement area. The simplest method is to use a mobile phone that displays the Evolved Absolute Radio Frequency Channel Number (EARFCN) as parameter (either by running a built-in code or accessing advanced networks configuration from the menu) [24]. However, this is device and OS-specific, and the exact steps to access the information might change in time because of updates. Another possibility is to use srsLTE and search for cells in the area of the experiment using `cell_search`:

```
$ ./ cell_search -b <freq_band>
```

`Cell_search` receives as parameter a frequency band `<freq_band>`, scans it and displays all the discovered cells operating in the input band. The frequency band allocated to operators is public. For example, the frequency band in the university area is available online on the Norwegian National Telecommunication Authority [28]. Mapping the scanned cells to operators can be easily done by a lookup in the frequency allocation reports of the European commission [12]. Of course, the same resource can be used to determine the frequency band too. Alternative solutions include intercepting and decoding System Information Block (SIB) messages, which offer additional information such as the priority of selection in case of multiple overlapping cells [25].

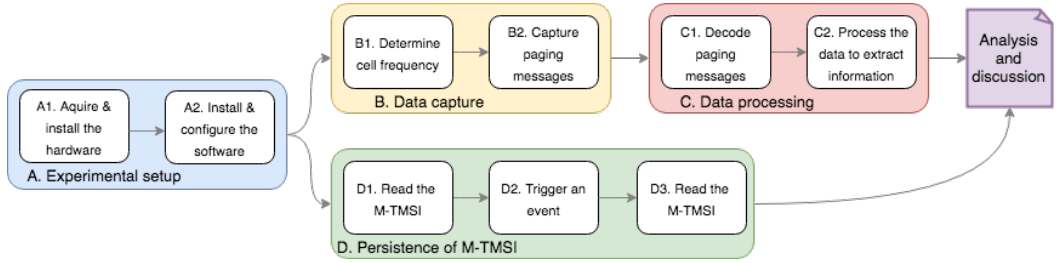

**Fig. 5** Schematic methodology applied for this work

### B2. Capture paging messages

Once the targeted downlink frequency `<dl_freq>` is known, the srsLTE prebuild application `pdsch_ue` can be used to capture paging messages:

```
$ ./ pdsch_ue -f <dl_freq> -r fffe
```

The eavesdropped cell is automatically selected in the given downlink frequency. When there are two or more cells running on the same frequency in the capture area, the cell with the stronger signal strength is selected. The value `fffe` is a fixed value for Paging-Radio Network Temporary Identifier (P-RNTI), which identifies paging and system information change notification [7].

### C. Data processing

After the paging messages are captured, the data has to be decoded to a human easily readable format, which further permits automatically extraction of information and analysis.

### C1. Decode paging messages

Decoding of paging messages is performed in two steps. First, a C code embedded in srsLTE decodes the paging message from Physical Downlink Shared Channel (PDSCH) to Abstract Syntax Notation One (ASN.1) hexadecimal format. This step is in fact performed directly by the tool during the message capture. Then, the ASN.1 format is translated to readable Extensible Markup Language (XML). Figure 6 illustrates a decoding of ASN.1 to XML using a free online tool [21]. However, for bulk decoding, we wrote a customised Python script that uses the external library *libmich* [22] and further used it in our experiments.

### C2. Process the data to extract information

Once in XML format, the data can be easily analysed. For example, a simple search on M-TMSI would reveal the presence of the temporary identity in the capture file. Simple customised scripts must be used in this phase to extract the information of interest from captures.

### D. Persistence of M-TMSI

To test the persistence of the temporary identifier M-TMSI, a different methodology is applied. This is because there is no need to capture and decode paging messages. The steps to conduct the experiment follow the lower branch (A-D) in Figure 5. In the experimental setup phase, the UE is connected to the commercial network. The M-TMSI is then read before and after an event is triggered or after a pre-defined period of time.

The M-TMSI is given by the last 4 bytes of the EPS Location Information field. The field is accessible on

#### Decode from direct input

1. Select an ASN.1 interface (*):
   RRC (PCCH-Message) 8.20.0

2. Copy below your hexadecimal encoding (max. 100 KB). Only the first PDU will be decoded.

```
# Below is an example for RRC DL-DCCH ASN.1 interface
# Decode this buffer,
# or choose your own ASN.1 interface with your own hexadecimal encoding
#
40038D03 F7390000
```

Decode buffer

(a) ASN.1 hexadecimal encoding

```
<PCCH-Message>
    <message>
        <c1>
            <paging>
                <pagingRecordList>
                    <PagingRecord>
                        <ue-Identity>
                            <s-TMSI>
                                <mmec>00111000</mmec>
                                <m-TMSI>11010000000111110111001110010000</m-TMSI>
                            </s-TMSI>
                        </ue-Identity>
                        <cn-Domain>
                            <ps/>
                        </cn-Domain>
                    </PagingRecord>
                </pagingRecordList>
            </paging>
        </c1>
    </message>
</PCCH-Message>

7 bytes decoded.
*** DECODING SUCCESSFUL ***
```

(b) XML format after decoding

**Fig. 6** Paging message decoded using an online tool [21]

**Fig. 7** M-TMSI read from the EPS Location Information on iPhone, iOS version 10

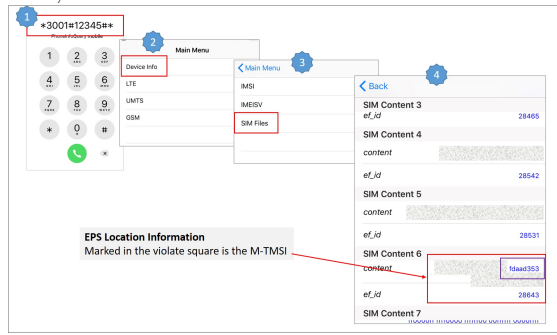

iPhone with iOS version 10, by dialling *3001#12345#* and identified by *ef_id* 28643 [6] . Figure 3.3 illustrates step-by-step how to find the M-TMSI. A more complex alternative to find the M-TMSI is to trigger paging messages to the UE, capture, decode them and read the M-TMSI. This can be done by techniques similar to the ones given by Shaik et al. [33].

## 4 Results and Discussion

Following the method described in Section 3, we conducted an experimental study to analyse the subscribers' privacy for three mobile operators in Norway: Telenor, Telia and ice.net. The experiments were performed in the wireless security lab at NTNU for educational and research purposes, and without disclosing any private information about commercial subscribers.

First, we analysed paging messages with respect to two criteria: type of paging messages (Section 4.1) and type of subscribers' identity contained in the paging messages (Section 4.2). Then, we tested the behaviour of a commercial subscriber in different scenarios with respect to identity protection by usage of temporary identifiers (Section 4.3). Last, we verified if smart paging is implemented in the operators' network (Section 4.4).

### 4.1 Paging Messages

All captures were performed during working hours, under similar conditions. The experiment has been performed several times, for different frequencies in the area of the wireless security lab at NTNU, with similar results.

For each operator, the number of captured paging records during three one-hour period samples are summarised in Table 1. In total, a significant number of records were captured from Telenor, the largest mobile operator in Norway and one of the world's largest mobile telecommunication company. Based on subscriptions in first half 2017, Telenor has 48,8% of the Norwegian market share, Telia 34,7%, ICE 6,4%, and others 10% [26]. We consider this to be a reason for the large number of paging records, in comparison to the other operators.

Most of the captured paging messages in this experiment contain paging records only. For Telenor, some `systemInfoModification` messages were captured too. It is not illustrated in Table 1, but expectedly, none of the messages had `etws-Indication` enabled. Although the notification messages are not of direct significance for the experiment, it is of interest to differentiate the messages that contain paging records and will be used for further analysis.

### 4.2 Identity of Subscribers

The captured messages were used to check the type of identity the network operators use for paging. For the conducted experiment, all paging records contain the temporary identifier M-TMSI. No IMSIs were found. This means that the operators successfully preserve the confidentiality of the permanent identity with respect to its presence in the paging messages. They are compliant to the standards, in the sense that they use temporary identifiers to protect the permanent identity of their subscribers.

For completeness, Table 2 contains the total number of paging messages, the number of paging messages that contain more than one paging record and the total number of paging records. Note that for all operators, there are paging messages with more than a single paging record. The percentage of messages that contain multiple identities is 12,17% for Telenor, followed by Telia with 7,12% and ice.net with 3,93%. As already noted, all identities used for paging are temporary M-TMSI.

### 4.3 Persistence of Temporary Identities

The usage of temporary identities is just a first step in protecting the privacy of the subscribers. If they are kept unchanged for a long time, their functionality is drastically reduced. This is because a subscriber can be easily identified and located by the temporary identity, without the necessity of knowing the permanent identity. So, changing the temporary identity often enough is a simple measure to mitigate attacks. This motivates the testing of the persistence of the temporary identities in different scenarios. More precise, we test under which circumstances the temporary identification is refreshed.

**Table 1**  Number of paging records, captured within one hour at different times of a week

| Network operator | Cell PHYID, frequency | Time | Number of paging records | Number of syst.modif. |
|---|---|---|---|---|
| Telenor | 112 816.0 MHz | 15.Jan.2018, Mon 08:43:19 - 09:43:19 | 65058 | - |
| | | 18.Jan.2018, Thu 16:54:32 - 17:54:32 | 43642 | - |
| | | 19.Jan.2018, Fri 10:22:45 - 11:22:45 | 72743 | 115 |
| Telia | 124 806.0 MHz | 15.Jan.2018, Mon 09:49:35 - 10:49:35 | 27138 | - |
| | | 18.Jan.2018, Thu 15:32:59 - 16:32:50 | 23485 | - |
| | | 19.Jan.2018, Fri 08:54:34 - 09:54:34 | 26332 | - |
| ice.net | 78 796.0 MHz | 26.Apr.2018, Thu 11:00:00 - 11:59:59 | 2356 | - |
| | | 27.Apr.2018, Fri 08:35:00 - 09:34:59 | 2205 | - |
| | | 30.Apr.2018, Mon 08:13:00 - 09:12:59 | 1840 | - |

**Table 2**  Number of paging records with M-TMSIs

| Network Operator | Number of paging messages | Number of paging messages with multiple M-TMSIs | Number of paging records with M-TMSIs |
|---|---|---|---|
| Telenor | 160480 | 19533 | 181443 |
| Telia | 71616 | 5101 | 76955 |
| ice.net | 6154 | 242 | 6401 |

**Table 3**  Verification of M-TMSI refreshment

| Event | M-TMSI refreshed (Yes/No/Uncertain) | |
|---|---|---|
| | Telia | ice.net |
| Mobile device switched off/on | Yes | Yes |
| Mobile flight mode on/off | Yes | Yes |
| Mobile originating voice call | Yes | No |
| Mobile terminating voice call | Yes | Yes |
| Mobile originating short message | No | No |
| Mobile terminating short message | No | No |
| Mobile originating data | No | No |
| Mobile terminating data | No | No |
| TA Change | Yes | Yes |
| Periodic TAU | Uncertain (no refresh in 48 hours) | Yes (refresh in less than 24h) |

Table 3 shows the results. There are no results for Telenor because the M-TMSI could not be read as explained in Subsection 3.3. As expected, switching the mobile off and on, as well as turning the flight mode on and off trigger the M-TMSI update. This is because the mobile needs to re-attach to the network. Similarly, voice calls (mobile originating and/or mobile terminating voice call) and moving from one TA to another trigger M-TMSI reallocation. Events such as mobile origi-nating and mobile terminating short messages or data do not trigger refreshment of the M-TMSI. This is expectable, for example in LTE the frequency of data transmission is high and this would trigger too much activity in the network with no benefit for security. Finally, we tested if the M-TMSI is refreshed when the UE camps on the same cell for several hours, with no call activity. In this case, the refresh should be done periodically by a special procedure, at a timer expiration. Expectedly, for ice.net the M-TMSI changed at least three times in 24 hours to plausibly random values. Sur-prisingly, for Telia the M-TMSI remained unchanged in a 48-hours time interval. How often a M-TMSI changes in time is an operator specific decision, being a tradeoff between security and efficiency. However, keeping the temporary identity unchanged for such long periods can threaten the privacy of the subscribers.

**Table 4** Smart paging analysis for Telenor

| Neighbour Cells | Cell 112 (816 MHz) | Cell 298 (1830 MHz) |
|---|---|---|
| Time Range | 19.01.2018 15:32:26 - 16:09:19 | 19.01.2018 15:29:58 - 16:07:29 |
| Number of paging records | 22752 | 8259 |
| Number of paging records in both cells at the same time | 1078 | |
| Distant Cells (≈ 2.2km) in the same TA | Cell 243 (816 MHz) | Cell 106 (816 MHz) |
| Time Range | 27.01.2018 16:15:00 - 17:15:59 | 27.01.2018 16:15:00 - 17:15:59 |
| Number of paging records | 77318 | 39935 |
| Number of paging records in both cells at the same time | 0 | |

**Table 5** Smart paging analysis for Telia

| Neighbour Cells | Cell 124 (806 MHz) | Cell 123 (806 MHz) |
|---|---|---|
| Time Range | 20.01.2018 10:37:25 - 11:19:30 | 20.01.2018 10:38:57 - 11:19:24 |
| Number of paging records | 16648 | 3713 |
| Number of paging records in both cells at the same time | 2811 | |
| Distant Cells (≈ 2.2km) in the same TA | Cell 124 (806 MHz) | Cell 280 (806 MHz) |
| Time Range | 27.01.2018 19:37:00 - 20:37:59 | 27.01.2018 19:37:00 - 20:37:59 |
| Number of paging records | 33675 | 33946 |
| Number of paging records in both cells at the same time | 0 | |

### 4.4 Smart Paging

To verify if smart paging is implemented, the paging messages are captured simultaneously in at least two cells within the same TA. Following the methodology, the captured paging messages are decoded, then the subscribers' identities and the timestamps are extracted and compared. For this experiment, we updated the srsLTE source code to timestamp the captured paging messages because the comparison should be done for the same timestamp. Without smart paging, the paging records should be identical because the paging is broadcasted within the whole TA. With smart paging, the paging records should be different because the paging is only broadcasted in the latest cell the UE camped on.

To achieve our goal, paging messages from two neighbour cells, respectively paging messages from distant cells (approximately 2.2km apart) but same TA, were captured and analysed. The downlink frequencies remain the same as in Table 1: 816 MHz for Telenor and 806 MHz for Telia. Figure 8, adapted from Finnsenderen [28], shows the geographical position of the capture and the targeted cells for Telia and Telenor. The two neighbour cells are both located in NTNU Gløshaugen campus. The distant cells are located in Moholt East.

Tables 4 and 5 illustrate the results for Telenor and Telia respectively. Each table contains the total number of paging records in each cell, and the number of records that page the same UE and are broadcasted in both cells simultaneously. The experiment result assumes that the same S-TMSI is for the same UE during the capturing period. For both Telia and Telenor, in the case of the neighbour cells, the amount of records in one cell is significantly less than in the other. This suggest that one cell is used as a backup for the other. There is an amount of records paging the same UE in the neighbour cells at the same time, but there is no simultaneously paging to the same UE in the distant cells, even if they are in the same TA. This means that both Telenor and Telia use smart paging. This increases the performance of the network by decreasing the number of broadcasted messages, but in the same time increases the risk of precise localisation of subscribers: a paged UE can be located within the coverage of one cell, and not at TA level.

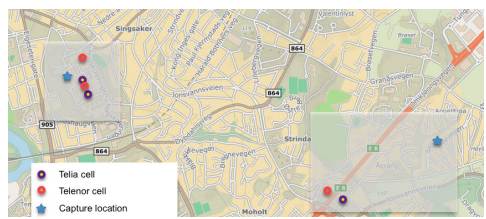

**Fig. 8** Locations of operator's cells and message capture

Independently of our experiment, for which only simultaneously broadcast of paging messages is of interest, there are some significant observations for the discussion. There were captured paging records for a UE in one cell only, and later simultaneously pages records for the same UE in both cells. This means that the UE did not reply to the paging message in the first cell, so it was paged on a broader area to increase the probability to reach the target. There are also cases where after paging a UE in one cell, a paging record for the same UE was captured to the second cell only. This indicates the movement of the UE from the first cell to the second, raising up possibilities for an adversary to track the movement of the UE in time, as long as the temporary identifier used for paging remains unchanged.

## 5 Conclusions

We described an experimental method to test the subscribers' privacy exposure by capturing and analysing LTE paging messages. We verified by experiment that paging messages can be captured and decoded by using only publicly available tools and minimal technical skills. In terms of identities, we describe experiments to test if mobile operators bring to minimum the usage of the permanent identity of subscribers by adopting and frequently updating the temporary identity. In terms of localisation, we describe a method to verify if smart paging is implemented. Smart paging increases performance by decreasing the number of paging messages in the network, but in the same time it brings the risks of a more precise localisation of subscribers. All experiments can be easily reproduced.

Following the experimental method described, we performed a case study on three LTE mobile operators in Norway. The results show that all operators protect the privacy of the permanent identity by not exposing it in paging messages. Moreover, our results show that the temporary identity update is triggered by specific events, in conformity to the 3GPP recommendations. However, the same does not hold for all operators when referred to the periodically update. Our experiments indicate that there is an operator that does not refresh the temporary identity in a 48 hours period.

To conclude, work must be done to reduce unprotected pre-authentication traffic to minimum in the next generation of mobile networks. Data capture is in general hard to detect and mitigate, so mobile operators must assure they implement the correct mechanisms to mitigate attacks.